\documentstyle[12pt,aps]{revtex}
\begin{document}
\title{Bethe ansatz solution of the anisotropic correlated electron model
associated with the Temperley-Lieb algebra }
\author{ A. Lima--Santos,$^1$ Itzhak Roditi$^{2}$ and Angela Foerster$^{3}$ }

\address{$^1$Departamento de Fisica, Universidade Federal de
S\~ ao Carlos, Caixa Postal 676,\\ 13569-905 S\~ ao Carlos, Brazil }

\address{$^2$Centro Brasileiro de
Pesquisas F\'{\i}sicas - CBPF,
Rua Dr. Xavier Sigaud 150, 22290-180, Rio de Janeiro, RJ - Brazil }

\address{ $^3$Instituto de F\'{\i}sica da UFRGS, Av. Bento Gon\c{c}alves
9500, Porto Alegre, RS - Brazil }

\maketitle
\begin{abstract}
\vspace{0.35cm}

A recently proposed strongly correlated electron system associated to the
Temperley Lieb
algebra is solved by means of the coordinate Bethe ansatz for periodic and
closed boundary conditions.
\end{abstract}
\vspace{0.35cm}

\section{Introduction}

Integrable highly correlated electron systems have been attracting
increasing interest due to their potential applications in condensed matter
physics. The prototypical examples of such systems are the Hubbard and t-J
models as well as their supersymmetric generalizations \cite{um}.
Recently many other correlated electron models have been formulated 
\cite{{dois},{doisa},{doisb},{doisc},{doisd},{tres},{tresa},{tresb}}.
Among these an interesting subclass corresponds to
models associated to the Temperley-Lieb (TL) algebras \cite{TLieb}. 
For such models there exists a well established method to construct a 
series of spin Hamiltonians as representations of the TL algebra and 
of quantum groups, the R matrix associated with the XXZ chain being 
the simplest example \cite{batch1}.
Later on, this approach was generalized by Zhang \cite{zhng} to 
construct graded representations of the TL algebra using Lie 
superalgebras and quantum supergroups. Along these lines a new 
isotropic strongly correlated electron model was obtained \cite{quatro}, 
as well as its anisotropic version with periodic and closed boundary 
conditions \cite{cinco}. In addition, it was shown in \cite{cinco} 
that this last choice of boundary generates a quantum group invariant
model, in contrast to the traditional periodic one.

Models with quantum group invariance and closed boundary conditions 
were first discussed by Martin \cite{martin} from representations 
of the Hecke algebra. More recently, by means of a generalized
algebraic Bethe ansatz, Karowski and Zapletal \cite{KZ} 
presented a class of quantum group invariant n-state vertex models
with closed boundary conditions. 
Within the framework of the coordinate Bethe ansatz closed 
spin chains invariant under $U_q(s\ell (2))$  were 
investigated by Grosse et al. \cite{grosse}.
Also, an extension of the algebraic approach to the case of 
graded vertex models \cite{skly} was analysed in \cite{foercb} 
where a $U_q(sp\ell (2,1))$
invariant susy t-J model with closed boundary conditions was presented.

In this paper we obtain through the coordinate Bethe ansatz approach the
solution of the anisotropic, or q-deformed, electronic model proposed in
\cite{cinco}, for periodic and closed boundary conditions. Here,
the meaning of closed is that an operator, coupling the first
and last sites, is introduced into the expression of the Hamiltonian such 
that we obtain a quantum algebra invariant closed system 
(see \cite{seis} for more details).
In particular, for the closed case the Bethe ansatz
equations are derived by extending the systematic procedure 
recently developed
in \cite{sete} to solve the quantum group invariant closed spin 1 chain
associated to the TL algebra for the case of a graded vertex model.

The paper is organized as follows. In section 2 we describe the correlated
electron system associated with the TL algebra. In section 3 we find
through the coordinate Bethe ansatz the
spectra of the model with usual periodic boundary conditions. In
section 4 the Bethe ansatz solution is presented for closed
boundary conditions. A summary of our main results is presented
in section 5.

\section{The Model}

The starting point for building the model is a 4-dimensional module V of the
Lie superalgebra $U_q(g\ell (2/1))$ utilized to obtain a representation of
the TL algebra. Let $\{|x \rangle \}^4_{x=1}$ be an orthonormal basis of
V
which carries the following parity,

\begin{equation}
[|1\rangle ]=[|4\rangle ]=0 \ \ \ \ \ [|2\rangle ]=[|3\rangle ]=1 .
\end{equation}
Everywhere we shall use the graded-tensor product law, defined by,

\[
(a\otimes b)(c\otimes d)=(-1)^{[b][c]} \ \ (ac\otimes bd)
\]
and also the rule,

\[
(|x\rangle \otimes |y\rangle )^{\dagger} ={(-1)^{[|x\rangle] [|y\rangle]
}}
\langle x|{\otimes} \langle y| .
\]

It is then possible to construct the following unnormalized vector of $V
\otimes V$

\begin{eqnarray}
\left| \Psi \right\rangle &=&q^{-1/2}\left| 4\right\rangle \otimes \left|
1\right\rangle +q^{1/2}\left| 1\right\rangle \otimes \left| 4\right\rangle
+q^{-1/2}\left| 3\right\rangle \otimes \left| 2\right\rangle -q^{1/2}\left|
2\right\rangle \otimes \left| 3\right\rangle  \nonumber \\
\left\langle \Psi \right| &=&q^{-1/2}\left\langle 4\right| \otimes
\left\langle 1\right| +q^{1/2}\left\langle 1\right| \otimes \left\langle
4\right| +q^{-1/2}\left\langle 3\right| \otimes \left\langle 2\right|
-q^{1/2}\left\langle 2\right| \otimes \left\langle 3\right|  \nonumber \\
&&  \label{eq0.1}
\end{eqnarray}

Next, to arrive at a hermitian Hamiltonian we consider the operator

\[
T=\left| \Psi \right\rangle \left\langle \Psi \right|
\]

A straightforward calculation shows that
\begin{eqnarray}
T^2 = [2(q &+& q^{-1})]T \nonumber \\
(T\otimes I)(I\otimes T)(T\otimes I) &=& T\otimes I \\
(I\otimes T)(T\otimes I)(I\otimes T) &=& I\otimes T \ ,  \nonumber
\end{eqnarray}
such that $T$ provides a representation of the $TL$ algebra. 
Following the approach of ref. \cite{batch1} to obtain solutions of 
the Yang-Baxter equation through the TL algebra, a 
local Hamiltonian can be defined by
(see ref. \cite{cinco} for more details).
\begin{equation}
H_{i,i+1}=T_{i,i+1},
\end{equation}
where on the N-fold tensor product space we denoted
\begin{equation}
T_{i,i+1}=I^{\otimes (i-1)}\otimes T\otimes I^{\otimes (N-i-1)}.
\end{equation}

In view of the grading the basis vectors of the module $V$ can
be identified with the eletronic states as follows
\[
|1\rangle \equiv |+-\rangle = c^+_+c^+_-|0\rangle \ ,
\ |2\rangle \equiv |-\rangle =c^+_-|0\rangle \ , \
|3\rangle \equiv |+\rangle = c^+_+|0\rangle \ ,
\ |4\rangle \equiv | 0\rangle
\]
allowing $H_{i,i+1}$ to be expressed as
\begin{eqnarray}
H_{i,i+1} &=& qn_{i,+}n_{i,-}(1-n_{i+1,+})
(1-n_{i+1,-})+q^{-1}(1-n_{i,+})(1-n_{i,-})n_{i+1,+}n_{i+1,-}
\nonumber \\
&+& q^{-1}n_{i,+}(1-n_{i,-})n_{i+1,-}(1-n_{i+1,+})+
qn_{i,-}(1-n_{i,+})n_{i+1,+}(1-n_{i+1,-})\nonumber \\
&-& S^+_iS^-_{i+1}-S^-_iS^+_{i+1}+
c^+_{i,+}c^+_{i,-}c_{i+1,-}c_{i+1,+}+
c^+_{i+1,+}c^+_{i+1,-}c_{i-}c_{i+}
\\
&+&qc^+_{i,+}c_{i+1,+}n_{i,-}(1-n_{i+1,-}) + h.c.
-c^+_{i,-}c_{i+1,-}n_{i,+}(1-n_{i+1,+}) + h.c.
\nonumber \\
&+& c_{i,-}c^+_{i+1,-}n_{i+1,+}(1-n_{i,+}) + h.c.
-q^{-1} c_{i,+}c^+_{i+1,+}n_{i+1,-}(1-n_{i,-})+h.c.\nonumber
\end{eqnarray}
where the $c^{(+)}_{i\pm}$ are spin up or down annihilation (creation)
operators, the $S'_is$ spin matrices and the $n'_is$ occupation
numbers of electrons at lattice site $i$.
This model describes electron pair hopping,
correlated hopping and  generalized spin interactions. In the
limit $q\rightarrow 1$ it reduces to the isotropic Hamiltonian 
introduced by Links in \cite{quatro}, which was shown to be 
invariant with respect to $ gl(2) \otimes u(1)$.

The global Hamiltonian is given by

\begin{equation}
H=\sum_{k=1}^{N-1} \, H_{k,k+1} \, \, \, + \, \, \, b.t.  \label{eq1.1}
\end{equation}
where $b.t.$ denotes the boundary term. The usual imposition of 
periodic boundary conditions (PBC),i.e., $ b.t. = H_{N,1}$, has 
the effect of breaking the $U_q(gl(2))\otimes u(1)$ symmetry of the 
model, since $H_{N,1} \neq H_{1,N}$, reflecting the non-cocommutativity 
of the co-product. However,  it was shown in \cite{cinco} 
following \cite{{martin},{KZ},{grosse},{foercb}} that for a special choice 
of the boundary term it is in fact possible to recover 
a quantum algebra invariant Hamiltonian which is in addition 
periodic in a certain sense. We shall call this type of boundary 
by closed one (CBC) and denote it by $b.t. = {\cal U}_{0} $
(see section 4 for details).

In the next sections we will find the spectrum of this Hamiltonian
for these two types of boundaries (PBC and CBC) through a modified
version of the coordinate Bethe ansatz.

\section{Bethe ansatz solution for periodic boundary conditions}

The case with periodic boundary conditions described by the Hamiltonian
\begin{equation}
H=\sum_{k=1}^{N} \, T_{k,k+1},   \label{eqpbc}
\end{equation}
can be mapped into a quantum spin
chain of $N$ sites each with spin $3/2$. To verify this we notice that 
the local Hamiltoniancan be rewritten as

\[
T=\left| \Psi \right\rangle \left\langle \Psi \right| =\left(
\begin{array}{ll}
U & {\huge 0}_{4\times 12} \\
{\huge 0}_{12\times 4} & {\Huge 0}_{12\times 12}
\end{array}
\right)
\]

where

\begin{equation}
U=
\begin{array}{l}
\left\langle 14\right| \\
\left\langle 23\right| \\
\left\langle 32\right| \\
\left\langle 41\right|
\end{array}
\left(
\begin{array}{c}
\begin{array}{llll}
\left| 14\right\rangle & 
\left| 23\right\rangle & 
\left| 32\right\rangle &
\left| 41\right\rangle
\end{array}
\\
\!\!\!\!
\begin{array}{llll}
q \, \, \,   & -q \, \, \, \, \, \,   & \, \, \, 1 \, \,\,\, \, \,  & \, \, \, 1 \\
-q & \, \, \, q\, & -1\, & -1 \\
1 & -1 & q^{-1} & q^{-1} \\
1 & -1 & q^{-1} & q^{-1}
\end{array}
\end{array}
\right)  \label{eq0.2}
\end{equation}
and the following correspondence has to be understood
 \begin{equation}
 \left| 1\right\rangle =\left(
 \begin{array}{l}
 1 \\
 0 \\
 0 \\
 0
 \end{array}
 \right) ,\left| 2\right\rangle =\left(
 \begin{array}{l}
 0 \\
 1 \\
 0 \\
 0
 \end{array}
 \right) ,\left| 3\right\rangle =\left(
 \begin{array}{l}
 0 \\
 0 \\
 1 \\
 0
 \end{array}
 \right) ,\left| 4\right\rangle =\left(
 \begin{array}{l}
 0 \\
 0 \\
 0 \\
 1
 \end{array}
 \right) ,  \label{eq1.2}
 \end{equation}
 The spin values are given by the eigenvalues of the operator $S^{z}$%
 \[
 S^{z}=\left(
 \begin{array}{llll}
 3/2 &  &  &  \\
 & 1/2 &  &  \\
 &  & -1/2 &  \\
 &  &  & -3/2
 \end{array}
 \right)
 \]
 and the total spin operator commuting with H is $S_{T}^{z}$
 \begin{equation}
 S_{T}^{z}=\sum_{k=1}^{N}{I^{\otimes (k-1)}}\otimes S^{z}\otimes
 {I^{\otimes (N-k)}}.
 \end{equation}
 Following \cite{sete} the spectrum of the above Hamiltonian can be
 classified in sectors which are defined by the eigenvalues of the operator
 number
 \begin{equation}
 r=\frac{3}{2}N-S_{T}^{z}  \label{eq1.3}
 \end{equation}

Let us now start to diagonalize $H$ in every sector, i.e., 
\[
H{\cal \ }\Psi =E\Psi 
\]

In the first sectors $r=0, 1, 2$ the eigenstates do not move under 
the action of $H$, i.e., $ H{\cal \ }\Psi_{r=0, 1, 2} = 0$. For this 
reason they are called impurities \cite{rol}.
In sector $r=3$, we
encounter the situation where the states $\left| \alpha ,k\right\rangle $
and $\left| -\alpha ,k\pm 1\right\rangle $, $\alpha \in S^z=\{\frac{-3}{2},%
\frac{-1}{2},\frac{1}{2},\frac{3}{2}\}$, occur in neighboring pairs. They
move under the action of $H$, {\it i.e.,} the sector $r=3$ contains one
free
{\em pseudoparticle}. In general, for a sector $r$ we may have $p$
pseudoparticles and $N_{\frac{1}{2}}$ and $N_{\frac{-1}{2}}$, impurities of
the type $\frac{1}{2}$ and $\frac{-1}{2}$, respectively, such that
\begin{equation}
r=3p+N_{\frac{1}{2}}+2N_{\frac{-1}{2}}  \label{pba1}
\end{equation}

For the first nontrivial sector $r=3$, the correspondent eigenspace is spanned
by the states $|k(-\alpha ,\alpha )>=|\frac{3}{2}\ \frac{3}{2}\cdots 
\frac{3}{2} {-\alpha }\alpha \ \frac{3}{2}\cdots
\frac{3}{2}>$
, where $k=1,2,...,N-1\ $and{\rm \ }$\ \alpha \in S^z.$ 
We seek eigenstates of
$H$ which are linear combinations of these vectors. It is very convenient to
consider the linear combination
\begin{equation}
\left| \Omega (k)\right\rangle =\left| k(\frac{-3}{2},\frac{3}{2}
)\right\rangle +\left| k(\frac{-1}{2},\frac{1}{2})\right\rangle -q\left| k(%
\frac{1}{2},\frac{-1}{2})\right\rangle +q\left| k(\frac{3}{2},\frac{-3}{2}%
)\right\rangle   \label{pba2}
\end{equation}
which is an eigenstate of $U_{k}$ \footnote{ From now on we will adopt the 
convention that $ U_{k} \equiv U_{k,k+1} $ operates in a direct product 
of complex spaces at positions $k$ and $k+1$}
\begin{equation}
U_{k}\left| \Omega (k)\right\rangle =(Q+Q^{-1})\left| \Omega
(k)\right\rangle =2(q+q^{-1})\left| \Omega (k)\right\rangle \text{.}
\label{pba3}
\end{equation}
and also a highest weight state, i.e., $ S^+ \Psi = 0$.
Moreover, the action of $U_{k\pm 1}$ on $\left| \Omega (k)\right\rangle $ is
very simple
\begin{equation}
\begin{array}{lll}
U_{k\pm 1}\left| \Omega (k)\right\rangle =\left| \Omega (k\pm
1)\right\rangle  &  & U_{k}\left| \Omega (k\pm 1)\right\rangle =\left|
\Omega (k)\right\rangle  \\
&  &  \\
U_{k}\left| \Omega (m)\right\rangle =0 &  & k\neq \{m\pm 1,m\}
\end{array}
\label{pba4}
\end{equation}

It should be emphasized that the linear combination (\ref{pba2}) affords a
considerable simplification in the diagonalization of $H$ in comparsion 
with the traditional calculus employing the usual spin basis \cite{sete}.
In fact, we believe that this type of ansatz is quite general and could 
be applied to solve a larger class of Hamiltonians derived from 
representations of the TL algebra.

We will now start to diagonalize $H$ in the sector $r=3$. Let us consider 
the non-trivial case of one
free pseudoparticle 

\begin{equation}
\Psi _{3}=\sum_{k}A(k)\left| \Omega (k)\right\rangle .  \label{pba5}
\end{equation}
Using the eigenvalue equation $H${\cal \ }$\Psi _{3}=E_{3}\Psi _{3}$, one
can derive a complete set of equations for the wavefunctions $A(k)$.

When the bulk of $H$ acts on $\left| \Omega (k)\right\rangle $ it sees the
reference configuration, except in the vicinity of $k$ where we use (\ref
{pba3}) and (\ref{pba4}) to get the following eigenvalue equation
\begin{eqnarray}
(E_{3}-Q-Q^{-1})A(k) &=&A(k-1)+A(k+1)  \nonumber \\
2 &\leq &k\leq N-2  \label{pba7}
\end{eqnarray}
Here we will treat periodic boundary conditions . They demand $%
U_{N,N+1}=U_{N,1}$, implying $A(k+N)=A(k)$. This permits us to complete 
the
set of equations (\ref{pba7}) for $A(k)$ by including the equations for $k=1$
and $k=N-1$. Now we parametrize $A(k)$ by plane wave $A(k)=A\xi ^{k}$ to 
get
the energy of one free pseudoparticle as:
\begin{eqnarray}
E_{3} &=&2(q+q^{-1})+\xi +\xi ^{-1}  \nonumber \\
\xi ^{N} &=&1  \label{pba8}
\end{eqnarray}
Here $\xi ={\rm e}^{i\theta }$, $\theta $ being the momenta determined from
the periodic boundary to be $\theta =2\pi l/N$, with $l$ an integer.

Let us consider the state with one pseudoparticle and one impurity of type $%
\frac{1}{2}$, which lies in the sector $r=4$. We seek eigenstates in the form

\begin{equation}
\Psi _{4}(\xi _{1},\xi _{2})=\sum_{k_{1}<k_{2}}\left\{
A_{1}(k_{1},k_{2})\left| \Omega _{1}(k_{1},k_{2})\right\rangle
+A_{2}(k_{1},k_{2})\left| \Omega _{2}(k_{1},k_{2})\right\rangle \right\}
\label{pba9}
\end{equation}
We try to build these eigenstates out of translationally invariant products
of one pseudoparticle excitation with parameter $\xi _{2}$ and one impurity
with parameter $\xi _{1}$:
\begin{equation}
\Psi _{4}(\xi _{1},\xi _{2})=\left| \frac{1}{2}(\xi _{1})\right\rangle
\times \Psi _{3}(\xi _{2})+\Psi _{3}(\xi _{2})\times \left| \frac{1}{2}(\xi
_{1})\right\rangle  \label{pba9a}
\end{equation}
Using one-pseudoparticle eigenstate solution (\ref{pba5}) and comparing this
with (\ref{pba9}) we get
\begin{eqnarray}
\left| \Omega _{1}(k_{1},k_{2})\right\rangle &=&\left| k_{1}(\frac{1}{2}%
),k_{2}(\frac{-3}{2},\frac{3}{2})\right\rangle +\left| k_{1}(\frac{1}{2}%
),k_{2}(\frac{-1}{2},\frac{1}{2})\right\rangle  \nonumber \\
&&-q\left| k_{1}(\frac{1}{2}),k_{2}(\frac{1}{2},\frac{-1}{2})\right\rangle
+q\left| k_{1}(\frac{1}{2}),k_{2}(\frac{3}{2},\frac{-3}{2})\right\rangle
\nonumber \\
&&  \nonumber \\
\left| \Omega _{2}(k_{1},k_{2})\right\rangle &=&\left| k_{1}(\frac{-3}{2},%
\frac{3}{2}),k_{2}(\frac{1}{2})\right\rangle +\left| k_{1}(\frac{-1}{2},%
\frac{1}{2}),k_{2}(\frac{1}{2})\right\rangle  \nonumber \\
&&-q\left| k_{1}(\frac{1}{2},\frac{-1}{2}),k_{2}(\frac{1}{2})\right\rangle
+q\left| k_{1}(\frac{3}{2},\frac{-3}{2}),k_{2}(\frac{1}{2})\right\rangle
\nonumber \\
&&  \label{pba10}
\end{eqnarray}
and
\begin{equation}
A_{1}(k_{1},k_{2})=A_{1}\xi _{1}^{k_{1}}\xi _{2}^{k_{2}}\qquad ,\qquad
A_{2}(k_{1},k_{2})=A_{2}\xi _{2}^{k_{1}}\xi _{1}^{k_{2}}.  \label{pba12}
\end{equation}
Periodic boundary conditions $A_{1}(k_{2},N+k_{1})=A_{2}(k_{1},k_{2})$ and $%
A_{i}(N+k_{1},N+k_{2})=A_{i}(k_{1},k_{2})$,\quad $i=1,2$ imply that
\begin{equation}
A_{1}\xi _{2}^{N}=A_{2}\quad ,\quad \xi ^{N}=(\xi _{1}\xi _{2})^{N}=1\
\label{pba13}
\end{equation}

When $H$ now acts on $\Psi _{4}$, we will get a set of coupled equations for
$A_{i}(k_{1},k_{2}),$ $i=1,2$. We split the equations into {\em far}
equations, when the pseudoparticle do not meet the impurity and {\em near}
equations, containing terms when they are neighbors.

Since the impurity is annihilated by $H$, the action of $H$ on (\ref{pba9})
in the case {\em far} ({\it i.e}., $(k_{2}-k_{1})\geq 3$), can be writen down
directly from (\ref{pba7}) :

\begin{equation}
\left( E_{4}-Q-Q^{-1}\right)
A_{1}(k_{1},k_{2})=A_{1}(k_{1},k_{2}-1)+A_{1}(k_{1},k_{2}+1)  \label{pba15}
\end{equation}
and similar equations for $A_{2}(k_{1},k_{2})$. Using the parametrization (%
\ref{pba12}), these equations will give us the energy eigenvalues
\begin{equation}
E_{4}=Q+Q^{-1}+\xi _{2}+\xi _{2}^{-1}  \label{pba16}
\end{equation}
To find $\xi _{2}$ we must consider the{\em \ near} equations. First, we
compute the action of $H$ on the coupled {\em near} states $\left| \Omega
_{1}(k,k+1)\right\rangle $ and $\left| \Omega _{2}(k,k+2)\right\rangle $:

\begin{eqnarray}
H\left| \Omega _{1}(k,k+1)\right\rangle  &=&(Q+Q^{-1})\left| \Omega
_{1}(k,k+1)\right\rangle +\left| \Omega _{1}(k,k+2)\right\rangle -\left|
\Omega _{2}(k,k+2)\right\rangle   \nonumber \\
&&  \label{pba17}
\end{eqnarray}
The last terms in these equations tell us that a pseudoparticle can
propagate past the isolated impurity, but in doing so causes a shift in its
position by two lattice sites. Substituting (\ref{pba17}) into the eigenvalue
equation, we get

\begin{equation}
\left( E_{4}-Q-Q^{-1}\right) A_{1}(k,k+1)=A_{1}(k,k+2)-A_{2}(k,k+2)
\label{pba20}
\end{equation}
These equations, which are not automatically satisfied by the ansatz (\ref
{pba12}), are equivalent to the conditions
\begin{equation}
A_{1}(k,k) = -A_{2}(k,k+2)\qquad   \label{pba21}
\end{equation}
obtained by subtracting Eq. (\ref{pba20}) from Eq.(\ref{pba15}) for $k_{1}=k$
$,$ $k_{2}=k+1$. The conditions (\ref{pba21}) require a modification of the
amplitude relation (\ref{pba13}):
\begin{equation}
\frac{A_{2}}{A_{1}}=-\xi _{1}^{-2}=\xi _{2}^{N}\Rightarrow \xi _{2}^{N}\xi
_{1}^{2}=-1\qquad \text{{\rm or}\qquad }\xi _{2}^{N-2}\xi ^{2}=-1
\label{pba22}
\end{equation}

In the sectors $3<r<6$ we also will find states, which consist of one
pseudoparticle with parameter $\xi _{r-2}$ interacting with $r-3$
impurities, distributing according to (\ref{pba1}), with parameters $\xi
_{i},i=1,2...,r-3$.

The energy of these states is parametrized as in (\ref{pba16}) and $\xi
_{r-2}$ satisfies the condition (\ref{pba22}) with $\xi =\xi _{1}\cdots \xi
_{r-3}\ \xi _{r-2}$. It involves only $\xi _{r-2}$ and $\xi _{{\rm imp}}=\xi
_{1}\ \xi _{2}\cdots \xi _{r-3}$, being therefore highly degenerate, {\it %
i.e.}
\begin{equation}
\xi _{r-2}^{N}\xi _{1}^{2}\ \xi _{2}^{2}\cdots \xi _{r-3}^{2}=(-1)^{r-3}
\label{pba24}
\end{equation}
This is to be expected due to the irrelevance of the relative distances, up
to jumps of two positions via exchange with a pseudoparticle. Moreover,
these results do not depend on the impurity type.

The sector $r=6$ contains, in addition to the cases discussed above, states
which consist of two interacting pseudoparticles. We seek eigenstates in the 
form
\begin{equation}
\Psi _{6}(\xi _{1},\xi _{2})=\sum_{k_{1}+1<k_{2}}A(k_{1},k_{2})\left| \Omega
(k_{1},k_{2})\right\rangle   \label{pba25}
\end{equation}

Applying $H$ to the state of (\ref{pba25}), we obtain a set of equations for
the wavefunctions $A(k_{1},k_{2})$. When the two pseudoparticles are
separated, ($k_{2}-k_{1}\geq 3$) these are the following {\em far}
equations:
\begin{equation}
\begin{array}{lll}
\left( E_{6}-2Q-2Q^{-1}\right) A(k_{1},k_{2}) & = &
A(k_{1}-1,k_{2})+A(k_{1}+1,k_{2}) \\
&  &  \\
& + & A(k_{1},k_{2}-1)+A(k_{1},k_{2}+1)
\end{array}
\label{pba30}
\end{equation}
We already know them to be satisfied, if we parametrize $A(k_{1},k_{2})$ by
plane waves (\ref{pba12}). The corresponding energy eigenvalue is
\begin{equation}
E_{6}=2Q+2Q^{-1}+\xi _{1}+\xi _{1}^{-1}+\xi _{2}+\xi _{2}^{-1}  \label{pbfa31}
\end{equation}

The real problem arises of course, when pseudoparticles are neighbors, so
that they interact and we have no guarantee that the total energy is a sum of
single pseudoparticle energies.

Acting of $H$ on the state $\left| \Omega (k,k+2)\right\rangle $ gives the
following set of equations for the {\em near} states
\begin{equation}
\begin{array}{lll}
H\left| \Omega (k,k+2)\right\rangle  & = & 2\left( Q+Q^{-1}\right) \left|
\Omega (k,k+2)\right\rangle +\left| \Omega (k-1,k+2)\right\rangle  \\
&  &  \\
& + & \left| \Omega (k,k+3)\right\rangle +U_{k+1}\left| \Omega
(k,k+2)\right\rangle
\end{array}
\label{pba32}
\end{equation}

Before we substitute this result into the eigenvalue equation, we observe
that some new states are appearing. In order to incorporate these new states
in the eigenvalue problem, we define
\begin{equation}
U_{k+1}\left| \Omega (k,k+2)\right\rangle = \left| \Omega
(k,k+1)\right\rangle +\left| \Omega (k+1,k+2)\right\rangle   \label{pba33}
\end{equation}
Here we underline that we are using the same notation for these new states.
Applying $H$ to them we obtain
\begin{equation}
\begin{array}{lll}
H\left| \Omega (k,k+1)\right\rangle  & = & \left( Q+Q^{-1}\right) \left|
\Omega (k,k+1)\right\rangle +\left| \Omega (k-1,k+1)\right\rangle  \\
&  &  \\
& + & \left| \Omega (k,k+2)\right\rangle
\end{array}
\label{pba34}
\end{equation}
Now, we extend (\ref{pba25}), the definition of $\Psi _{6}$ , to
\begin{equation}
\Psi _{6}(\xi _{1},\xi _{2})=\sum_{k_{1}<k_{2}}A(k_{1},k_{2})\left| \Omega
(k_{1},k_{2})\right\rangle   \label{pba35}
\end{equation}
Substituting (\ref{pba32}) and (\ref{pba34}) into the eigenvalue equation,
we obtain the following set of {\em near} equations
\begin{equation}
\left( E_{6}-Q-Q^{-1}\right) A(k,k+1)=A(k-1,k+1)+A(k,k+2)  \label{pba36}
\end{equation}
Using the same plane wave parametrization for these new wavefunctions, the
equation (\ref{pba36}) gives us the {\em phase shift} produced by the
interchange of the two interacting pseudoparticles
\begin{equation}
\frac{A_{21}}{A_{12}}=-\frac{1+\xi +(Q+Q^{-1})\xi _{2}}{1+\xi +(Q+Q^{-1})=
\xi
_{1}}  \label{pba37}
\end{equation}
We thus arrive to the  (BAE) which fix the values of $\xi
_{1}$ and $\xi _{2}$ in the energy equation (\ref{pbfa31})
\begin{eqnarray}
\xi _{2}^{N} &=&-\frac{1+\xi +(Q+Q^{-1})\xi _{2}}{1+\xi +(Q+Q^{-1})\xi _{=
1}}
\nonumber \\
\xi ^{N} &=&(\xi _{1}\xi _{2})^{N}=1  \label{pba38}
\end{eqnarray}

In a generic sector $r$ with $l$ impurities parametrized by $\xi _{1}\xi
_{2}\cdots \xi _{l}$ and $p$ pseudoparticles with parameters $\xi _{l+1}\xi
_{l+2}\cdots \xi _{l+p}$, the energy is
\begin{equation}
E_{r}=\sum_{n=l+1}^{p}\left\{ Q+Q^{-1}+\xi _{n}+\xi _{n}^{-1}\right\}
\label{pba59}
\end{equation}
with $\xi _{n}$ determined by the Bethe ansatz equations
\begin{eqnarray}
\xi _{a}^{N}\xi _{1}^{2}\xi _{2}^{2}\cdots \xi _{l}^{2}
&=&(-1)^{l}\prod_{b\neq a=l+1}^{p}\left\{ -\frac{1+\xi _{b}\xi
_{a}+(Q+Q^{-1})\xi _{a}}{1+\xi _{a}\xi _{b}+(Q+Q^{-1})\xi _{b}}\right\} 
\nonumber \\
a &=&l+1,l+2,...,l+p,\qquad p\geq 2  \nonumber \\
\xi _{l+1}^{N}\xi _{1}^{2}\xi _{2}^{2}\cdots \xi _{l}^{2} &=&(-1)^{l},\qquad
p=1  \nonumber \\
\xi _{c}^{N-2p} &=&(-1)^{p},\quad c=1,2,...,l  \nonumber \\
\xi ^{N} &=&1,\qquad \xi =\xi _{1}\xi _{2}\cdots \xi _{l}\xi _{l+1}\xi
_{l+2}\cdots \xi _{l+p}.  \label{pba60}
\end{eqnarray}
The energy eigenvalues and the Bethe equations depend on the deformation
parameter $q$, through the relation $Q+Q^{-1}=2q+2q^{-1}$.

\section{Bethe ansatz solution for closed boundary conditions}

The quantum group invariant closed TL Hamiltonians which can be written
as \cite{cinco}:

\begin{equation}
H=\sum_{k=1}^{N-1}U_{k}+{\cal U}_{0}  \label{cbah}
\end{equation}
where $U_{k}$ is a Temperley-Lieb operator and ${\cal U}_{0}$ is a boundary
term defined through of a operator $G$ which plays the role of the
translation operator
\begin{equation}
{\cal U}_{0}=GU_{N-1}G^{-1}\quad ,\quad G=(Q-U_{1})(Q-U_{2})\cdots
(Q-U_{N-1})  \label{cba1}
\end{equation}
satisfying $[H,G]=0$ and additionally invariance with respect to the
quantum
algebra. The operator $G$ shifts the $U_{k}$ by one unit $%
GU_{k}G^{-1}=U_{k+1}$ and maps ${\cal U}_{0}$ into $U_{1}$, which manifest
the translational invariance of $H$. In this sense the Hamiltonian (\ref{cbah}) is 
periodic. From the physical point of view, this type of models
exhibit behavior similar to closed chains with twisted boundary conditions
(see for example \cite{pal2} for the case of the XXZ chain).

The action of the operator $G$ on the states $\left| \Omega (k)\right\rangle
$ can be easily computed using (\ref{pba2}), (\ref{pba3}) and 
(\ref{pba4}): It is simple on the bulk and at
the left boundary

\begin{equation}
G\left| \Omega (k)\right\rangle =-Q^{N-2}\ \left| \Omega (k+1)\right\rangle
{\rm \quad },{\rm \quad }1\leq k\leq N-2  \label{cba5}
\end{equation}
but has a non-trivial contribution at the right boundary

\begin{equation}
G\left| \Omega (N-1)\right\rangle =Q^{N-2}\sum_{k=1}^{N-1}(-Q)^{-k}\ \left|
\Omega (N-k)\right\rangle   \label{cba6}
\end{equation}
Similarly, the action of the operator $G^{-1}=(Q^{-1}-U_{N-1})\cdots
(Q^{-1}-U_{1})$ is simple on the bulk and at the right boundary

\begin{equation}
G^{-1}\left| \Omega (k)\right\rangle =-Q^{-N+2}\left| \Omega
(k-1)\right\rangle \ \quad ,\quad 2\leq k\leq N-1  \label{cba7}
\end{equation}
and non-trivial at the left boundary
\begin{equation}
G^{-1}\left| \Omega (1)\right\rangle =Q^{-N+2}\sum_{k=1}^{N-1}(-Q)^{k}\
\left| \Omega (k)\right\rangle .  \label{cba8}
\end{equation}

Now we proceed the diagonalization of $H$ as was made for the periodic case.
As (\ref{cbah}) and (\ref{eq1.1}) have the same bulk,{\it \ i.e.},
differences arise from the boundary terms only, we will keep all 
results relating
to the bulk of the periodic case presented in the previous section.

Let us consider one free pseudoparticle which lies in the sector $r=3$

\begin{equation}
\Psi _{3}=\sum_{k=1}^{N-1}A(k)\left| \Omega (k)\right\rangle .  
\label{cba9}
\end{equation}
The action of the operator ${\cal U}=\sum_{k=1}^{N-1}U_{k}$ on the states $%
\left| \Omega (k)\right\rangle $ is:

\begin{eqnarray}
{\cal U}\left| \Omega (1)\right\rangle  &=&(Q+Q^{-1})\left| \Omega
(1)\right\rangle +\left| \Omega (2)\right\rangle   \nonumber \\
{\cal U}\left| \Omega (k)\right\rangle  &=&(Q+Q^{-1})\left| \Omega
(k)\right\rangle +\left| \Omega (k-1)\right\rangle +\left| \Omega
(k+1)\right\rangle   \nonumber \\
\qquad \quad \text{{\rm for }\ }2 &\leq &k\leq N-2  \nonumber \\
{\cal U}\left| \Omega (N-1)\right\rangle  &=&(Q+Q^{-1})\left| \Omega
(N-1)\right\rangle +\left| \Omega (N-2)\right\rangle .  \label{cba10}
\end{eqnarray}
and using (\ref{cba5})--(\ref{cba8}) one can see that the action of ${\cal U}%
_{0}=GU_{N-1}G^{-1}$ vanishes on the bulk
\begin{equation}
{\cal U}_{0}\left| \Omega (k)\right\rangle =0\quad ,\quad 2\leq k\leq N-2
\label{cba13}
\end{equation}
and has the following contributions at the boundaries
\begin{equation}
{\cal U}_{0}\left| \Omega (1)\right\rangle =-\sum_{k=1}^{N-1}\ (-Q)^{k}\
\left| \Omega (k)\right\rangle ,\quad {\cal U}_{0}\left| \Omega
(N-1)\right\rangle =-\sum_{k=1}^{N-1}\ (-Q)^{-N+k}\ \left| \Omega
(k)\right\rangle .  \label{cba14}
\end{equation}
which are connected by
\begin{equation}
{\cal U}_{0}\left| \Omega (N-1)\right\rangle =(-Q)^{-N}\ {\cal U}_{0}\left|
\Omega (1)\right\rangle .  \label{cba15}
\end{equation}

Before we substitute these results into the eigenvalue equation, we will
define two new states

\begin{equation}
\left| \Omega (0)\right\rangle ={\cal U}_{0}\left| \Omega (1)\right\rangle
,\quad \ \left| \Omega (N)\right\rangle ={\cal U}_{0}\left| \Omega
(N-1)\right\rangle   \label{cba16}
\end{equation}
to include the cases $k=0$ and $k=N$ into the definition of $\Psi _{3}$,
equation (\ref{cba9}). Finally, the action of $H={\cal U}+{\cal U}_{0}$ on
the states $\left| \Omega (k)\right\rangle $ is

\begin{eqnarray}
H\left| \Omega (0)\right\rangle  &=&(Q+Q^{-1})\left| \Omega (0)\right\rangle
+(-Q)^{N}\left| \Omega (N-1)\right\rangle +\left| \Omega (1)\right\rangle 
\nonumber \\
&&  \nonumber \\
H\left| \Omega (N)\right\rangle  &=&(Q+Q^{-1})\left| \Omega (N)\right\rangle
+\left| \Omega (N-1)\right\rangle +(-Q)^{-N}\left| \Omega (1)\right\rangle
\label{cba17}
\end{eqnarray}
Substituting these results into the eigenvalue equation\ $H\Psi _{3}=E_{3}\
\Psi _{3}$ we get a complete set of eigenvalue equations for the
wavefunctions

\begin{eqnarray}
E_{3}\ A(k) &=&(Q+Q^{-1})A(k)+A(k-1)+A(k+1)  \nonumber \\
\quad \qquad \text{{\rm for }}1 &\leq &k\leq N-1  \label{cba18}
\end{eqnarray}
provided the following boundary conditions
\begin{equation}
(-Q)^{N}A(k)=A(N+k)\text{ }  \label{cba20}
\end{equation}
are satisfied.

The plane wave parametrization $A(k)=A\xi ^{k}$ solves these eigenvalue
equations and the boundary conditions provided that:
\begin{eqnarray}
E_{3} &=&Q+Q^{-1}+\xi +\xi ^{-1}\quad   \nonumber \\
\xi ^{N} &=&(-Q)^{N}  \label{cba21}
\end{eqnarray}
where $\xi ={\rm e}^{i\theta }$ and $\theta $ being the momentum.

Let us now consider the sector $r=6$, where we can find an eigenstate with
two interacting pseudoparticles. We seek the corresponding eigenfunction as
products of single pseudoparticles eigenfunctions, {\it i.e}.
\begin{equation}
\Psi _{6}=\sum_{k_{1}+1<k_{2}}A(k_{1},k_{2})\left| \Omega
(k_{1},k_{2})\right\rangle   \label{cba22}
\end{equation}

To solve the eigenvalue equation $H\Psi _{6}=E_{6}\Psi _{6}$, we recall
(\ref{pba4}) to get the action of ${\cal U}$ and ${\cal U}_{0}$ on the states $%
\left| \Omega (k_{1},k_{2})\right\rangle $. Here we have to consider four
cases: ({\it i})\ when the two pseudoparticles are separated in the bulk,
the action of ${\cal U}$ is
\begin{eqnarray}
{\cal U}\left| \Omega (k_{1},k_{2})\right\rangle  &=&2(Q+Q^{-1})\left|
\Omega (k_{1},k_{2})\right\rangle +\left| \Omega
(k_{1}-1,k_{2})\right\rangle +\left| \Omega (k_{1}+1,k_{2})\right\rangle 
\nonumber \\
&&+\left| \Omega (k_{1},k_{2}-1)\right\rangle +\left| \Omega
(k_{1},k_{2}+1)\right\rangle   \label{cba24}
\end{eqnarray}
i.e., for $k_{1}$ $\geq 2$ and $k_{1}+3\leq k_{2}\leq N-2$; ({\it ii}) when
the two pseudoparticles are separated but one of them or both are at the
boundaries
\begin{eqnarray}
{\cal U}\left| \Omega (1,k_{2})\right\rangle  &=&2(Q+Q^{-1})\left| \Omega
(1,k_{2})\right\rangle +\left| \Omega (2,k_{2})\right\rangle +\left| \Omega
(1,k_{2}-1)\right\rangle   \nonumber \\
&&+\left| \Omega (1,k_{2}+1)\right\rangle   \label{cba25}
\end{eqnarray}
\begin{eqnarray}
{\cal U}\left| \Omega (k_{1},N-1)\right\rangle  &=&2(Q+Q^{-1})\left| \Omega
(k_{1},N-1)\right\rangle +\left| \Omega (k_{1}-1,N-1)\right\rangle 
\nonumber \\
&&+\left| \Omega (k_{1}+1,N-1)\right\rangle +\left| \Omega
(k_{1},N-2)\right\rangle   \label{cba26}
\end{eqnarray}
\begin{equation}
{\cal U}\left| \Omega (1,N-1)\right\rangle =2(Q+Q^{-1})\left| \Omega
(1,N-1)\right\rangle +\left| \Omega (2,N-1)\right\rangle +\left| \Omega
(1,N-2)\right\rangle   \label{ba27}
\end{equation}
where $2\leq k_{1}\leq N-4$ and $4\leq k_{2}\leq N-2$; ({\it iii}) when the
two pseudoparticles are neighbors in the bulk
\begin{eqnarray}
{\cal U}\left| \Omega (k,k+2)\right\rangle  &=&2(Q+Q^{-1})\left| \Omega
(k,k+2)\right\rangle +\left| \Omega (k-1,k+2)\right\rangle +\left| \Omega
(k,k+3)\right\rangle   \nonumber \\
&&+U_{k+1}\left| \Omega (k,k+2)\right\rangle   \label{cba28}
\end{eqnarray}
for $2\leq k\leq N-4$ and ({\it iv}) when the two pseudoparticles are
neighbors and at the boundaries
\begin{equation}
{\cal U}\left| \Omega (1,3)\right\rangle =2(Q+Q^{-1})\left| \Omega
(1,3)\right\rangle +\left| \Omega (1,4)\right\rangle +U_{2}\left| \Omega
(1,3)\right\rangle   \label{cba29}
\end{equation}
\begin{eqnarray}
{\cal U}\left| \Omega (N-3,N-1)\right\rangle  &=&2(Q+Q^{-1})\left| \Omega
(N-3,N-1)\right\rangle +\left| \Omega (N-4,N-1)\right\rangle   \nonumber \\
&&+U_{N-2}\left| \Omega (N-3,N-1)\right\rangle   \label{cba30}
\end{eqnarray}

Moreover, the action of ${\cal U}_{0}$ does not depend on the
pseudoparticles are neither separated nor neighbors. It is vanishes in the
bulk
\begin{equation}
{\cal U}_{0}\left| \Omega (k_{1},k_{2})\right\rangle =0\quad \text{{\rm for}
\quad }k_{1}\neq 1\ \text{{\rm and} }k_{2}\neq N-1,  \label{cba31}
\end{equation}
and different of zero at the boundaries:
\begin{eqnarray}
{\cal U}_{0}\left| \Omega (1,k_{2})\right\rangle
&=&-\sum_{k=1}^{k_{2}-2}(-Q)^{k}\left| \Omega (k,k_{2})\right\rangle
-(-Q)^{k_{2}-1}U_{k_{2}}\left| \Omega (k_{2}-1,k_{2}+1)\right\rangle  \\
&&-\sum_{k=k_{2}+2}^{N-1}(-Q)^{k-2}\left| \Omega (k_{2},k)\right\rangle
\label{cba32}
\end{eqnarray}
\begin{equation}
{\cal U}_{0}\left| \Omega (k_{1},N-1)\right\rangle =(-Q)^{-N+2}\ {\cal U}%
_{0}\left| \Omega (1,k_{2})\right\rangle   \label{cba33}
\end{equation}
where $2\leq k_{1}\leq N-3$ and $3\leq k_{2}\leq N-2$.

Following the same procedure of one-pseudoparticle case we again define new
states in order to have consistency between bulk and boundaries terms
\begin{eqnarray}
{\cal U}_{0}\left| \Omega (1,k_{2})\right\rangle  &=&\left| \Omega
(0,k_{2})\right\rangle ,\quad {\cal U}_{0}\left| \Omega
(k_{1},N-1)\right\rangle =\left| \Omega (k_{1},N)\right\rangle   \nonumber \\
{\cal U}_{0}\left| \Omega (1,N-1)\right\rangle  &=&\left| \Omega
(0,N-1)\right\rangle +\left| \Omega (1,N)\right\rangle   \nonumber \\
U_{k+1}\left| \Omega (k,k+2)\right\rangle  &=&\left| \Omega
(k,k+1)\right\rangle +\left| \Omega (k+1,k+2)\right\rangle   \label{cba34}
\end{eqnarray}
Acting with $H$ on these new states, we get
\begin{eqnarray}
H\left| \Omega (0,k_{2})\right\rangle  &=&2(Q+Q^{-1})\left| \Omega
(0,k_{2})\right\rangle +\left| \Omega (0,k_{2}-1)\right\rangle +\left|
\Omega (0,k_{2}+1)\right\rangle   \nonumber \\
&&+\left| \Omega (1,k_{2})\right\rangle +(-Q)^{N-2}\left| \Omega
(k_{2},N-1)\right\rangle   \label{cba35}
\end{eqnarray}
\begin{eqnarray}
H\left| \Omega (k_{1},N)\right\rangle  &=&2(Q+Q^{-1})\left| \Omega
(k_{1},N)\right\rangle +\left| \Omega (k_{1}-1,N)\right\rangle +\left|
\Omega (k_{1}+1,N)\right\rangle   \nonumber \\
&&+\left| \Omega (k_{1},N-1)\right\rangle +(-Q)^{-N+2}\left| \Omega
(1,k_{1})\right\rangle   \label{cba36}
\end{eqnarray}
\begin{equation}
H\left| \Omega (k,k+1\right\rangle =(Q+Q^{-1})\left| \Omega
(k,k+1\right\rangle +\left| \Omega (k-1,k+1\right\rangle +\left| \Omega
(k,k+2\right\rangle   \label{cba37}
\end{equation}
Substituting these results into the eigenvalue equation, we get the
following equations for wavefunctions corresponding to the separated
pseudoparticles.
\begin{eqnarray}
(E_{6}-2Q-2Q^{-1})A(k_{1},k_{2}) &=&A(k_{1}-1,k_{2})+A(k_{1}+1,k_{2})
\nonumber \\
&&+A(k_{1},k_{2}-1)+A(k_{1},k_{2}+1)  \label{cba38}
\end{eqnarray}
{\it i.e}., for $k_{1}\geq 1$ and $k_{1}+3\leq k_{2}\leq N-1$. The boundary
conditions read now
\begin{equation}
A(k_{2},N+k_{1})=(-Q)^{N-2}A(k_{1},k_{2}).  \label{cba39}
\end{equation}
The parametrization for the wavefunctions

\begin{equation}
A(k_{1},k_{2})=A_{12}\xi _{1}^{k_{1}}\xi _{2}^{k_{2}}+A_{21}\xi
_{1}^{k_{2}}\xi _{2}^{k_{1}}  \label{cba40}
\end{equation}
solves the equation (\ref{cba38}) provided that
\begin{equation}
E_{6}=2(Q+Q^{-1})+\xi _{1}+\xi _{1}^{-1}+\xi _{2}+\xi _{2}^{-1}
\label{cba41}
\end{equation}
and the boundary conditions (\ref{cba39}) provided that
\begin{equation}
\xi _{2}^{N}=(-Q)^{N-2}\frac{A_{21}}{A_{12}}\quad ,\quad \xi
_{1}^{N}=(-Q)^{N-2}\frac{A_{12}}{A_{21}}\Rightarrow \xi ^{N}=(-Q)^{2(N-2)}
\label{cba42}
\end{equation}
where $\xi =\xi _{1}\xi _{2}=e^{i(\theta _{1}+\theta _{2})}$, $\theta
_{1}+\theta _{2}$ being the total momenta.

Now we include the new states (\ref{cba34}) into the definition of $\Psi _{6}
$ in order to extend (\ref{cba22}) to
\begin{equation}
\Psi _{6}=\sum_{k_{1}<k_{2}}A(k_{1},k_{2})\left| \Omega
(k_{1},k_{2}\right\rangle .  \label{cba43}
\end{equation}
Here we have used the same notation for separated and neighboring states.
Substituting (\ref{cba28}) and (\ref{cba37}) into the eigenvalue equation,
we get
\begin{equation}
(E_{6}-Q-Q^{-1})A(k,k+1)=A(k-1,k+1)+A(k,k+2)  \label{cba44}
\end{equation}
which gives us the phase shift produced by the interchange of the two
pseudoparticles
\begin{equation}
\frac{A_{21}}{A_{12}}=-\frac{1+\xi +(Q+Q^{-1})\xi _{2}}{1+\xi +(Q+Q^{-1})\xi
_{1}}.  \label{cba45}
\end{equation}
We thus arrive to the Bethe ansatz equations which fix the values of $\xi
_{1}$ and $\xi _{2}$:

\begin{eqnarray}
\xi _{2}^{N} &=&(-Q)^{N-2}\left\{ -\frac{1+\xi +(Q+Q^{-1})\xi _{2}}{1+\xi
+(Q+Q^{-1})\xi _{1}}\right\} ,  \nonumber \\
\quad \xi _{1}^{N}\xi _{2}^{N} &=&(-Q)^{2(N-2)}  \label{cba46}
\end{eqnarray}

Thus in the sector $r=3p$, we expect that the $p$-pseudoparticle phase shift
will be a sum of
two-pseudoparticle phase shifts and the
energy is given by
\begin{equation}
E_{3p}=\sum_{n=1}^{p}\left\{ Q+Q^{-1}+\xi _{n}+\xi _{n}^{-1}\right\}
\label{cba48}
\end{equation}
where
\[
\xi _{a}^{N}=(-Q)^{N-2p+2}\prod_{b\neq a}^{p}\left\{ -\frac{1+\xi _{a}\xi
_{b}+(Q+Q^{-1})\xi _{a}}{1+\xi _{a}\xi _{b}+(Q+Q^{-1})\xi _{b}}\right\}
,\quad a=1,...,p
\]
\begin{equation}
\left( \xi _{1}\xi _{2}\cdots \xi _{p}\right) ^{N}=(-Q)^{p(N-2p+2)}
\label{cba49}
\end{equation}
The corresponding eigenstates are
\begin{equation}
\Psi _{r}(\xi _{1},\xi _{2},...\xi _{p})=\sum_{1\leq k_{1}<...<k_{p}\leq
N-1}A(k_{1},k_{2,}...,k_{p})\left| \Omega
(k_{1},k_{2},...,k_{p})\right\rangle   \label{cba49a}
\end{equation}
where $\left| \Omega (k_{1},k_{2},...,k_{p})\right\rangle =\otimes
_{i=1}^{p}\left| \Omega (k_{i})\right\rangle $ and the wavefunctions satisfy
the following boundary conditions
\begin{equation}
A(k_{1},k_{2,}...,k_{p},N+k_{1})=(-Q)^{N-2p+2}A(k_{1},k_{2,}...,k_{p})
\label{cba49b}
\end{equation}

It is not all, in a sector $r$ we may have $p$ pseudoparticles and $N_{\frac{1%
}{2}},N_{\frac{-1}{2}}$ impurities of the type $\frac{1}{2},\frac{-1}{2}$,
respectively. Since $H$ is a sum of projectors on spin zero, these states
are also annihilated by ${\cal U}_{0}$ . Therefore the impurities play here
the same role as in the periodic case. It means that for a sector $r$ with
$l$ impurities with parameters $\xi _{1},...,\xi _{l}$ and $p$ pseudoparticles
with parameters $\xi _{l+1},...,\xi _{l+p}$ the energy is given by
(\ref{cba49}), and the Bethe equations do not depend on impurity type and are
given by
\begin{equation}
\xi _{a}^{N}\xi _{1}^{2}\xi _{2}^{2}\cdots \xi
_{l}^{2}=(-1)^{l}(-Q)^{N-2p+2} \prod_{b=l+1, b\neq a}^{l+p}
\left\{
-\frac{1+\xi _{a}\xi _{b}+(Q+Q^{-1})\xi _{a}}{1+\xi _{a}\xi
_{b}+(Q+Q^{-1})\xi _{b}}\right\}   \label{cba51}
\end{equation}
with $a=l+1,l+2,...,l+p\quad ,\quad p\geq 1$, and
\begin{equation}
\xi ^{2p}(\xi _{l+1}\cdots \xi _{l+p})^{N-2p}=(-1)^{l}(-Q)^{p(N-2p+2)}
\label{cba52}
\end{equation}
where $\xi =
\xi _{1}\xi _{2}\cdots \xi _{l}\xi _{l+1}\cdots \xi _{l+p}$.

Notice in the BAE (\ref{cba51}) the presence of a
special "$q$- term" $((-Q)^{N-2p+2})$ in comparsion with the
corresponding ones with usual periodic boundary conditions (\ref{pba60}).
In fact, this feature also appeared in other models
\cite{{KZ},{grosse},{foercb}}
and seems to be a peculiarity of quantum group invariant closed spin
chains.

\section{Conclusions}
We have applied the coordinate Bethe ansatz to find the
spectra of the anisotropic correlated electron system associated with
the TL algebra. This procedure was carried out for periodic and closed
boundary conditions and the differences between both cases
have been remarked.

We believe that the methods here presented could also be applied to
solve a larger class of Hamiltonians derived from representations
of the graded TL algebra, such as the orthosympletic models discussed
by Zhang in \cite{zhng}. This is presently under investigation.

Another  interesting extension of this work would be to adapt the methods
employed in this paper to solve multiparametric versions of these
models. \cite{{doze},{mult}}

\vspace{1cm}

{\bf Acknowledgment:}
The support of CNPq - Conselho Nacional de
Desenvolvimento Cient\'{\i}fico e Tecnol\'ogico is gratefully
acknowledged.
A.L.S also thanks FAPESP -
Funda\c{c}\~{a}o de Amparo \`{a} Pesquisa do Estado de S\~{a}o Paulo
for financial assistance. A.F. would like to thank the Institute
f\"ur Theoretische Physik - FUB for its kind hospitality, particularly
M. Karowski. She also thanks DAAD - Deutscher Akademischer Austauschdienst
and FAPERGS -  Funda\c{c}\~{a}o de Amparo \`{a} Pesquisa do Estado
do Rio Grande do Sul for financial support.

\end{document}